\documentclass[aps,prl,twocolumn,amsmath,amssymb,nofootinbib,superscriptaddress]{revtex4-2}
\usepackage{times}
\usepackage[pdftex]{graphicx}
\usepackage{dcolumn}
\usepackage{bm}
\usepackage{amsmath}
\usepackage{indentfirst}
\usepackage{float}
\usepackage[colorlinks]{hyperref}
\usepackage[dvipsnames]{xcolor}
\usepackage{xcolor}

\colorlet{shadecolor}{gray!40}

\usepackage{subfigure}
\usepackage{algorithm}
\usepackage{algorithmicx}
\usepackage{algpseudocode}
\usepackage{amsmath}
\usepackage{verbatim}
\usepackage{makecell}
\usepackage[normalem]{ulem}

\begin{document}

\title{Realization of an Error-Correcting Surface Code with Superconducting Qubits}

\author{Youwei Zhao}
\thanks{These three authors contributed equally}
\author{Yangsen Ye}
\thanks{These three authors contributed equally}
\author{He-Liang Huang}
\thanks{These three authors contributed equally}
\author{Yiming Zhang}
\author{Dachao Wu}
\author{Huijie Guan}
\author{Qingling Zhu}
\author{Zuolin Wei}
\author{Tan He}
\author{Sirui Cao}
\author{Fusheng Chen}
\author{Tung-Hsun Chung}
\author{Hui Deng}
\author{Daojin Fan}
\author{Ming Gong}
\author{Cheng Guo}
\author{Shaojun Guo}
\author{Lianchen Han}
\author{Na Li}
\author{Shaowei Li}
\author{Yuan Li}
\author{Futian Liang}
\author{Jin Lin}
\author{Haoran Qian}
\author{Hao Rong}
\author{Hong Su}
\author{Lihua Sun}
\author{Shiyu Wang}
\author{Yulin Wu}
\author{Yu Xu}
\author{Chong Ying}
\author{Jiale Yu}
\author{Chen Zha}
\author{Kaili Zhang}
\author{Yong-Heng Huo}
\author{Chao-Yang Lu}
\author{Cheng-Zhi Peng}
\author{Xiaobo Zhu}
\author{Jian-Wei Pan}
\affiliation{Department of Modern Physics, University of Science and Technology of China, Hefei 230026, China}
\affiliation{Shanghai Branch, CAS Center for Excellence in Quantum Information and Quantum Physics, University of Science and Technology of China, Shanghai 201315, China}
\affiliation{Shanghai Research Center for Quantum Sciences, Shanghai 201315, China}



\begin{abstract}
Quantum error correction is a critical technique for transitioning from noisy intermediate-scale quantum (NISQ) devices to fully fledged quantum computers. The surface code, which has a high threshold error rate, is the leading quantum error correction code for two-dimensional grid architecture. So far, the repeated error correction capability of the surface code has not been realized experimentally. Here, we experimentally implement an error-correcting surface code, the distance-3 surface code which consists of 17 qubits, on the \textit{Zuchongzhi} 2.1 superconducting quantum processor. By executing several consecutive error correction cycles, the logical error can be significantly reduced after applying corrections, achieving the repeated error correction of surface code for the first time. This experiment represents a fully functional instance of an error-correcting surface code, providing a key step on the path towards scalable fault-tolerant quantum computing.

\end{abstract}

\maketitle

\section{Introduction}
Progress towards scalable fault-tolerant quantum computation relies on exploiting quantum error correction (QEC) to protect quantum systems against inevitable noises~\cite{terhal2015quantum,lidar2013quantum,devitt2013quantum,bennett1996mixed,shor1995scheme,calderbank1996good,huang2020superconducting}. Notable experimental implementations on a variety of QEC architectures include surface code~\cite{andersen2020repeated,ai2021exponential,erhard2021entangling,marques2021logical}, repetition code~\cite{ai2021exponential}, Bosonic code~\cite{hu2019quantum,gertler2021protecting,ofek2016extending}, Shor code~\cite{luo2021quantum,egan2020fault}, color code~\cite{nigg2014quantum,ryan2021realization}, [[5,1,3]] code~\cite{gongexperimental} etc.~\cite{huang2021emulating,liu2019demonstration}. The surface code is unarguably the leading candidate for fault-tolerant quantum computation, featuring a high threshold error rate and requiring only nearest neighbor interactions on a 2-dimensional square lattice~\cite{kitaev2003fault,raussendorf2007fault,fowler2012surface}. This property makes it is perfectly compatible with the fabricate devices using planar photolithography, such as superconducting and quantum dot systems. In 2014, R. Barends $et$ $al.$ first realized high fidelity quantum gates at the fault-tolerant threshold for the surface code in a superconducting quantum processor~\cite{barends2014superconducting}. Surface code experiments have since been progressively developed in terms of scale and utility. The entangling operations between two 4-qubit surface codes using lattice surgery has been demonstrated in a ion-trap quantum processor~\cite{erhard2021entangling}. Most recently, the surface code with distance-two has been realized with seven qubits to demonstrate the repeated error detection~\cite{marques2021logical,andersen2020repeated,ai2021exponential,marques2021logical}.
Until now, all the surface code experiments have only the ability to detect errors, and repeated error correction of surface code has not been implemented in any experiment. However, as the system size and circuit depth grow, the fidelity decreases exponentially, making it impractical to rely solely on error detection and dropping error events.

The ability of repeated error correction is essential for realizing large-scale quantum algorithms, but it is significantly more difficult than just error detection. On the one hand, more redundant qubits are required for error correction than for error detection. However, scaling the number of qubits while maintaining high-fidelity quantum operations remains a key challenge for quantum computing. On the other hand, one needs to know the exact number of qubits that are corrupted and more importantly, their location in the quantum state and the type of error.

In this work, we present the first implementation of repeated error detection and correction of the surface code. Specifically, by encoding the logical state using a distance-three surface code on the \textit{Zuchongzhi} 2.1 superconducting quantum system~\cite{wu2021strong,zhu2021quantum}, we show that the logical error can be reduced by approximately $20\%$ after applying corrections in post-processing. We also test the error detection performance of this code, and observe that the lifetime of the logical qubit is longer than those of any constituent physical qubits, when we post-select the instances that no error is detected by both data qubits measurements and the stabilizer measurements in any cycle. Our studies, for the first time, demonstrate the feasibility of repeated quantum error correction using surface code, guiding future efforts to realize more powerful large-scale quantum error correction.

\begin{figure}[!htbp]
\begin{center}
\includegraphics[width=1\linewidth]{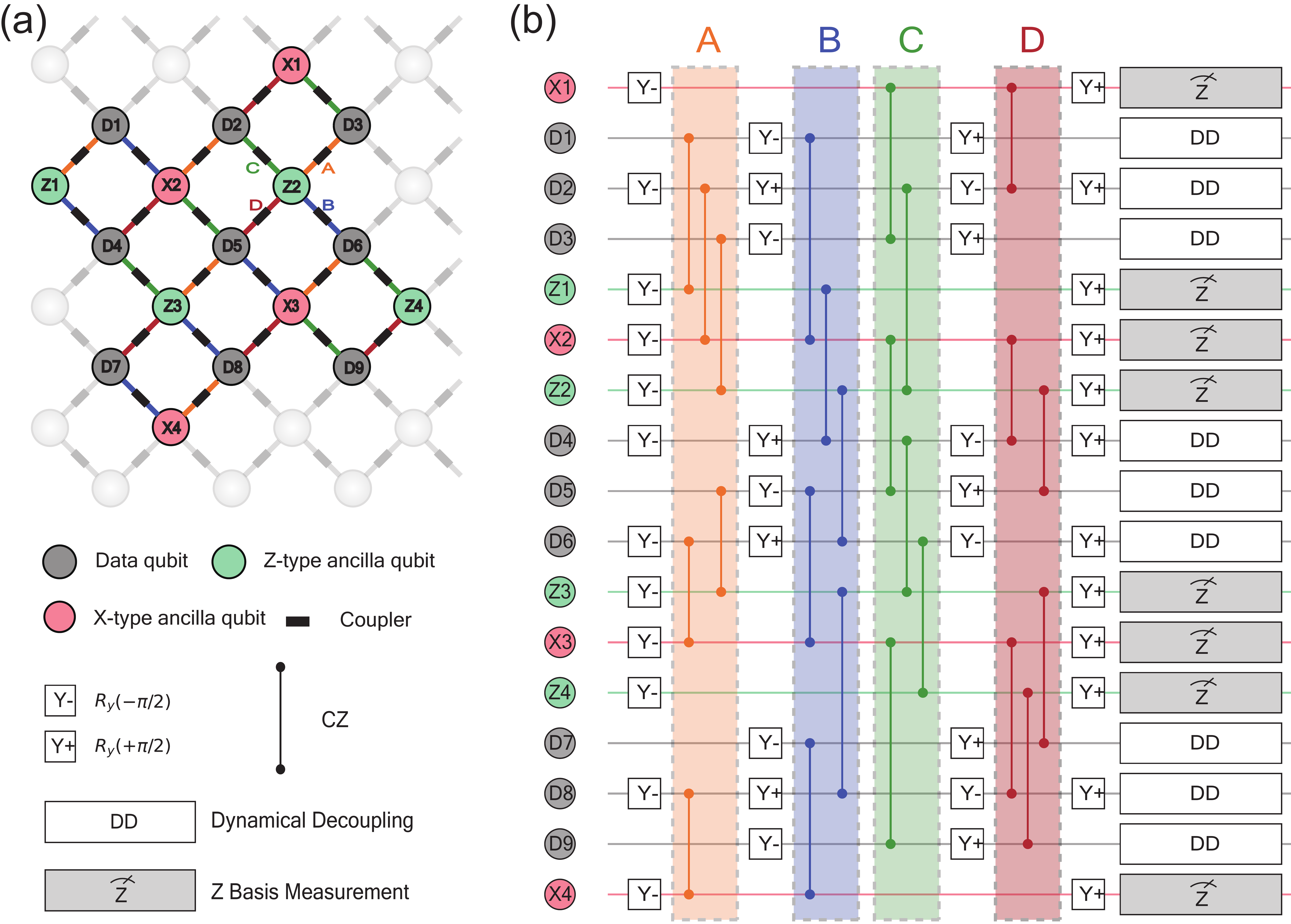}
\end{center}
\caption{\textbf{Layout and circuit implementation.} \textbf{(a)} Structure schematic of distance-three surface code. 17 qubits are choosen from \textit{Zuchongzhi} 2.1 superconducting quantum processor, with 8 data qubits(gray dots), 4 $Z$-type ancilla qubits(green dots) and 4 $X$-type ancilla qubits(red dots). Each pair of qubits are connected with a coupler(black rectangle). Connecting lines are colored according to their involvement in two-qubit gate layers as shown in (b). \textbf{(b)} Circuit for one error correction cycle. Dots on the left are in one-to-one correspondence to those in subfigure (a). Squares with $Y-$ and $Y+$ represents $Y$ rotation by an angle of $-\pi/2$ and $\pi/2$. Line with two dots denotes a controlled-phase gate (CZ). All gates in one color block are applied simultaneously. Gray rectangles with $Z$ denote a measurement in the Z basis. Block labelled DD are for dynamical decoupling operators.}
\label{fig1}
\end{figure}

\section{Encoding the Logical Qubit with a distance-3 surface code}
The surface code, as suggested by \cite{kitaev2003fault,raussendorf2007fault,fowler2012surface} can be implemented on a two-dimensional array of physical qubits with only nearest-neighbour coupling, making its realization readily available on the \textit{Zuchongzhi} 2.1 superconducting qubit platform. Here, we have chosen 17 out of the 66 qubits from the \textit{Zuchongzhi} 2.1 system and created a distance-three surface code.

Its structure are depicted in Fig.~\ref{fig1}(a). This 17-qubit surface code consists of 9 data qubits and 8 measurement qubits. The data qubits, which store the computation quantum state, are represented by gray dots with label $Dj$, $j = 1,2, \ldots, 9$. There are two types of measurement qubits. The green dots whose label starts with $Z$ describe Z-ancilla qubits which measure the Z-parity of its adjacent data qubits, while a red dot, whose label has an $X$ in it, depicts an X-ancilla qubit that checks the sign of the product of Pauli X operators acting on its neighboring data qubits. Taking measurement over all ancilla qubits projects the code space into the subspace spanned by the eigenstates of these parity check operators. The corresponding outcomes completely describe the state of the system. And any error occurred amid a process will manifest itself as a change in the measurement of these ancilla qubits. This enables us to keep track of the evolution of the system with only ancilla qubits without the risk of corrupting it.

Between each data qubit and measurement qubit, there resides a coupler denoted by a square rectangle. The introduction of couplers allows us to dynamically turn on and off interactions between nearest neighbour qubits~\cite{2014coupler}. On one hand, this allows us to implement entangling operations on adjacent pair of qubits while being protected from unwanted crosstalk. This makes it possible to pack all two-qubit gates in one surface code cycle into four layers. One the other, the architecture also alleviates the frequency crowding problems, making it easier to scale up the quantum system.

Figure~\ref{fig1}(b) illustrates the layout of circuit for one surface code cycle. Each row describes gate operations on one qubit, the order of which are set according to their rows in the physical configuration as depicted in Fig.~\ref{fig1}(a). The gates are applied from right to left in 9 time steps. All operations encapsulated in the same coloured box are applied simultaneously. The color of each two-qubit gate also matches that of the edge in Fig.~\ref{fig1}(a) to facilitate understanding of the order of gates in a physical context.

The line with two dots represents a controlled-phase gate, i.e. CZ gate. The circuit is modified from the one used in Ref.~\cite{versluis2017scalable} with all Hadamard gates replaced by a Y rotation, the latter of which are more native to our platform. The replacement is based on the fact that a Hadamard gate can be decomposed as $H = Z R_Y(-\pi/2)$ or $H= R_Y(\pi/2) Z$ and $(I\otimes Z)\, \text{CZ}\, (I\otimes Z) = \text{CZ}$.

As a last step of the surface code cycle, all states of the ancilla qubits are measured in the Z basis, where an extra Hadamard gate, or more apparently an extra Y gate, on the X-ancilla qubit produces the needed X measurement. In the meanwhile, a dynamical decoupling operation (DD)~\cite{ai2021exponential}~
is applied to the data target to mitigate dephasing problem of the data qubits. 


\section{System calibration}

\begin{figure}[tbp]
\begin{center}
\includegraphics[width=1.\linewidth]{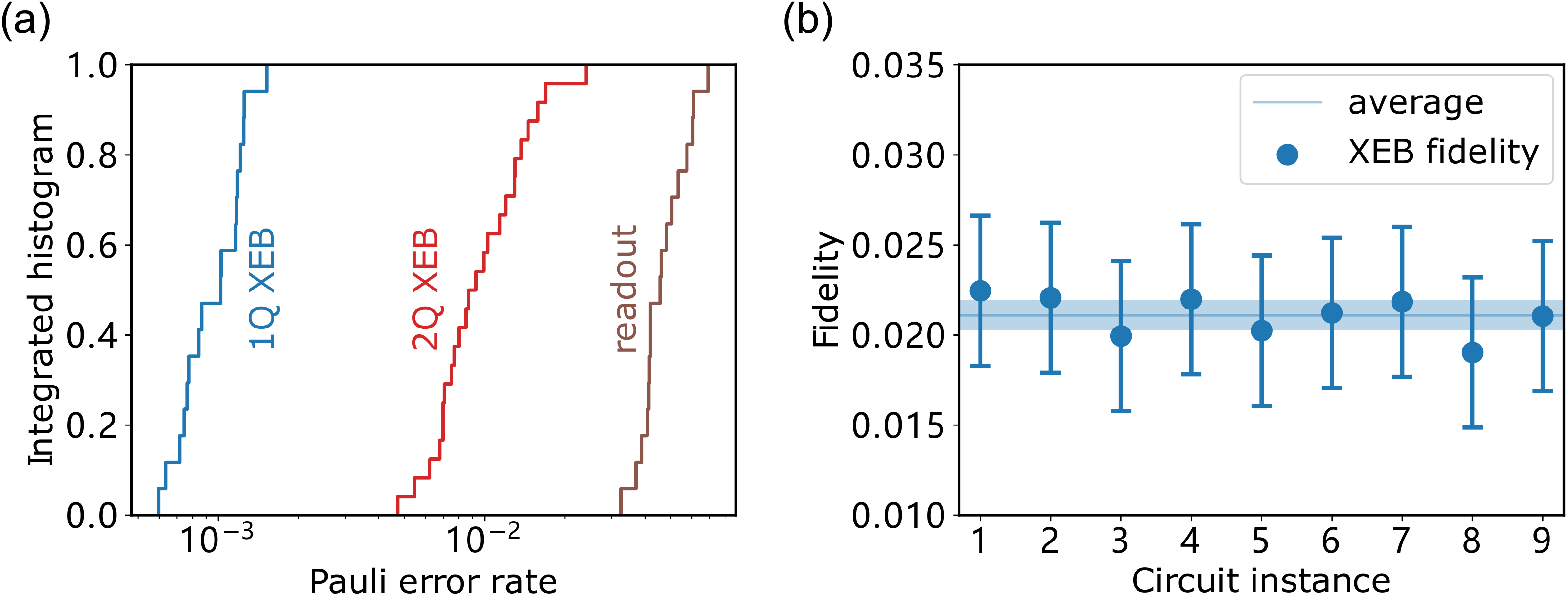}
\end{center}
\caption{\textbf{System calibration.} \textbf{(a)} Integrated histogram of single qubit $\pi/2$ rotation(1Q XEB), CZ gate(2Q XEB) and readout error(readout). 
\textbf{(b)} Cross entropy benchmarking fidelity from random sampling tasks for 9 instances of random circuits. Solid dots represents average fidelity over repetitions of sampling task with the same random seed. The average fidelity is 0.021. Error bars describes $\pm 5\sigma$ statistical deviation, color band is for $\pm 1\sigma$ with $\sigma= 1/\sqrt{N_{\text{sample}}}$, where $N_{\text{sample}}=1,440,000$. 
}
\label{fig_2_xeb}
\end{figure}

Figure~\ref{fig_2_xeb}(a) displays the integrated histogram for single-qubit gate, CZ gate and readout error after performing calibration. Each single-qubit gate can be implemented in 25ns with an average error of $0.098\%$. The duration of each CZ gates is set to be 32ns, with the average error being $1.035\%$. 
The measurement operation takes 1.5$\mu s$. With this setting, average readout error is $4.752\%$.
Since the readout readline width $\kappa/2\pi$ is large in our experimental setup, to reduce the effect of photon residue in the resonator before gate recommence, we conservatively insert an idle operation of 2.4 $\mu s$ after an measurement, extending the time cost to 3.9 $\mu s$ for an measurement operation.

Not only do we need to know how well we are doing with basic quantum operations, it is also necessary to assure ourselves that the complex state preparation and circuit operation are well understood. For this purpose,  we carried out a random circuit sampling task tailored for the surface code experiment, and compute the linear cross-entropy benchmarking fidelity, which is defined as
\begin{align}
{F_{\text{XEB}}} = {2^n}\langle P({x_i})\rangle  - 1,
\end{align}
to characterize the performance in a circuit level, where $n$ is the number of qubits, $P(x_i)$ represents the probability of bitstring $x_i$ computed for the ideal quantum circuit.

\begin{figure}[tbp]
\begin{center}
\includegraphics[width=1.0\linewidth]{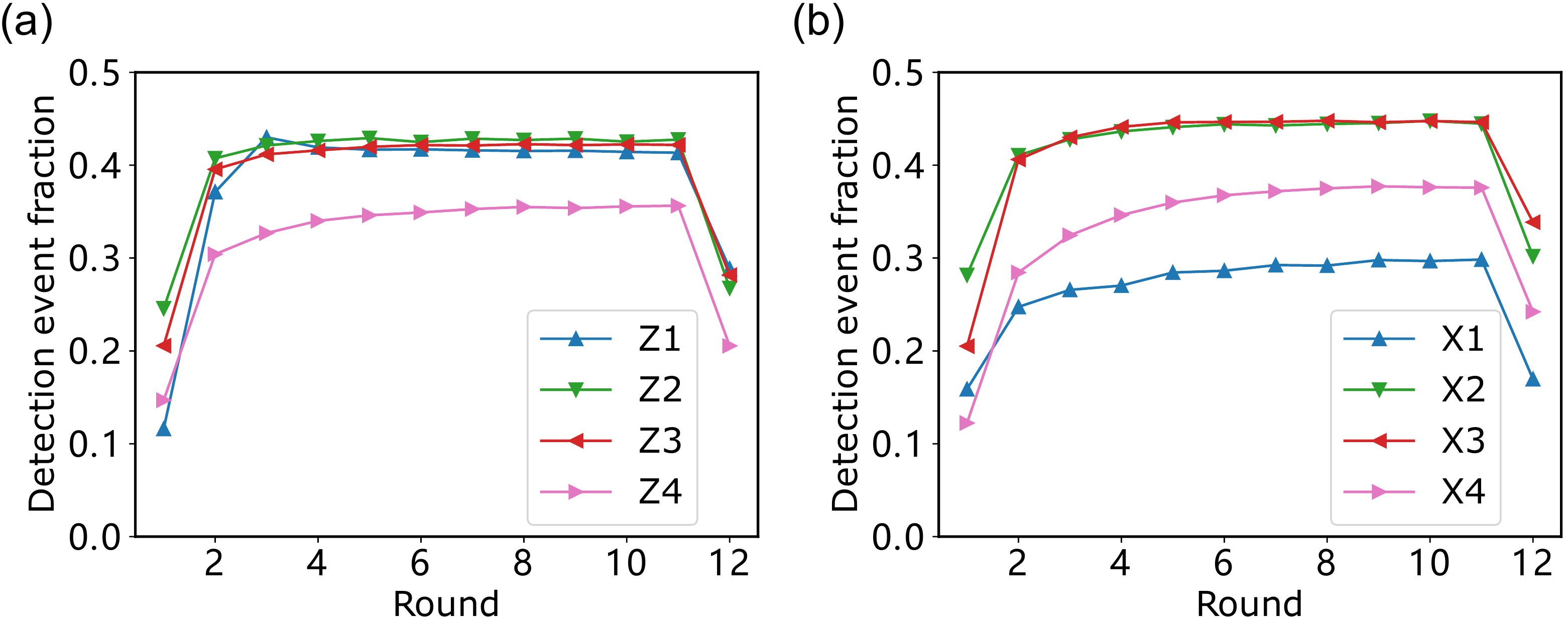}
\end{center}
\caption{\textbf{Error detection rate.} \textbf{(a)} Detection event fraction curve for logical $|0_L\rangle$ state as a function of surface code cycles. Various lines describe the fraction of samples with an error detected on the corresponding Z-type ancilla qubit. 
\textbf{(b)} Detection event fraction for logical $|-_L\rangle$ state.}
\label{fig_4_def}
\end{figure}

The random circuit is made up of alternating layers of two-qubit gate and single-qubit gate, but it is slightly different from the previous works~\cite{arute2019quantum,wu2021strong,zhu2021quantum}. To properly extract the effect of gate sequence to the fidelity, the two-qubit gates are chosen from the set \{A, B, C, D\} in the same order as the surface code cycle. The single-qubit gate is set in the same fashion as how the \textit{Zuchongzhi} 2.1 is benchmarked~\cite{wu2021strong,zhu2021quantum}. To be explicit, a random gate is selected from the pool of gates \{$R_X(\pi/2)$, $R_Y(\pi/2)$ and $R_{X+Y}(\pi/2)$\} and applied to each of the 17 qubits, with the only requirement such that subsequent single-qubit gate cannot be the same. In total, we stack 21 single-qubit gate layers and 20 two-qubit gate layers in the circuit for the sampling task.

We have sampled 9 instances of random circuits with different random seeds (see Fig.~\ref{fig_2_xeb}(b)),
and the average fidelity achieved is $0.021\pm0.001$. We also calculate a prediction of this value, which is 0.028, by taking the product of the Pauli fidelity of each single-qubit, two-qubit and measurment operation. As can be seen from the result, the fidelity of each gate is capable of predicting the circuit performance notwithstanding the inevitability of crosstalk and tailing in wavefonts.


\section{experimental results}

\begin{figure}[tbp]
\begin{center}
\includegraphics[width=1.0\linewidth]{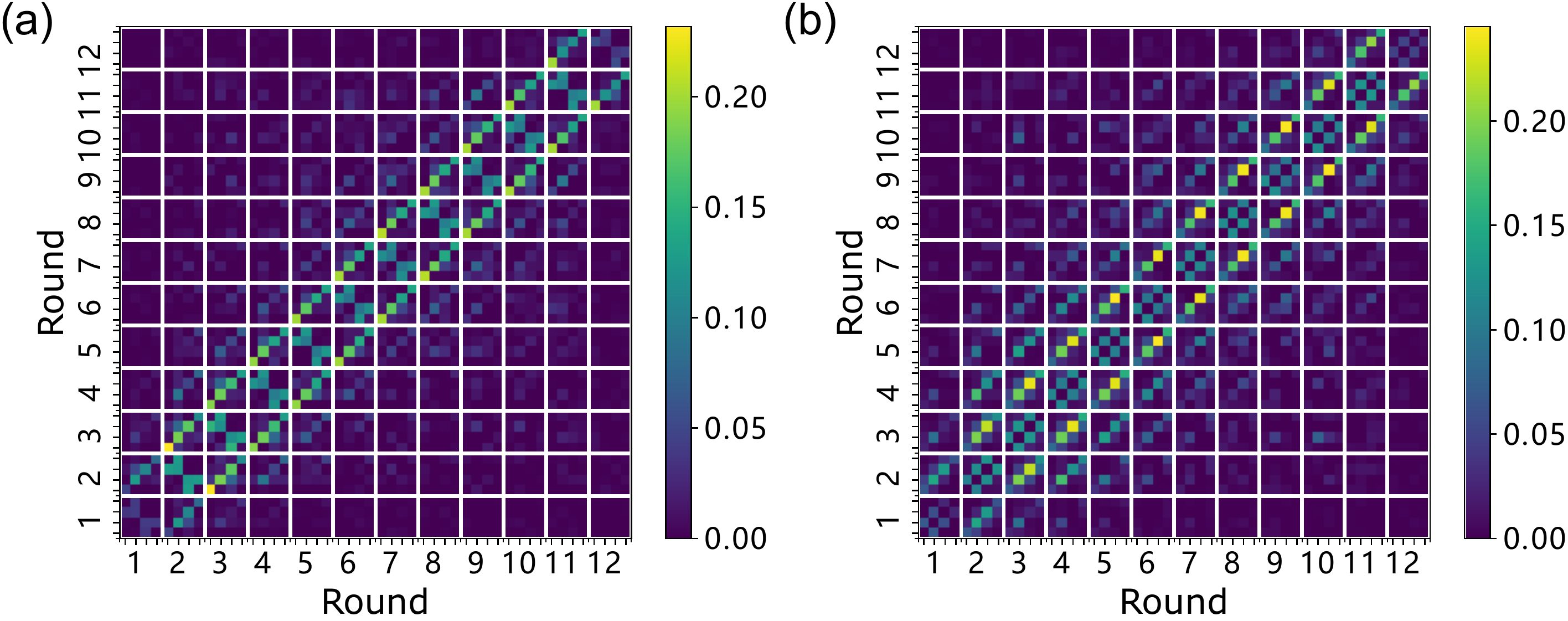}
\end{center}
\caption{\textbf{Error detection correlation.} \textbf{(a)} Correlation matrix for logical $|0_L\rangle$ state. The finer scale is used for the marking of ancilla qubits and each block has a definite cycle index.
Color scheme is presented in the side color bar with the dark side for low correlation and yellowish side for strong correlation. 
\textbf{(b)} Correlation matrix for logical $|-_L\rangle$ state.}
\label{fig_5_mat}
\end{figure}

With enough information about fidelity in both gate and circuit levels, we are ready to carry out the experiment. For the experiment, we repeat the surface code cycle as shown in Fig.~\ref{fig1}(b) up to 11 times. At the end of each stack, the states of the four measure qubits, either X or Z ancilla, are taken. For the last clock cycle, we also measure the state of the data qubits in the same basis as the syndrome measurement. This will give us an extra piece of information about the stabilizer. In total, for the circuit with $n$ cycles, we record $n$ sets of measurement about the ancilla qubits and one set of outcome for data qubit.

A direct information one can extract from these data is error occurrence. To do so, the state of a measure qubit takes an XOR operator with that of its previous run, giving rise to the value of a stabilizer relating to this cycle. This operation serves the same purpose as a reset gate acting on the ancilla qubits, which is missing in our circuit. The stabilizer on the first run simply adopts the value of the reading from the first measure. Aside from these stabilizer values, an extra record is created by comparing the calculated ancilla qubit state from the parity of the data qubits to the values measured in the last run. Whenever a changes occurs in subsequent stabilizer values, a detection event of errors is fired.

Making use of the 12 columns of stabilizer values, we can execute detection event analysis. The result is shown in Fig.~\ref{fig_4_def}. Figure~\ref{fig_4_def}(a) describes the fraction of a detection event (DEF) for the logical $|0_L\rangle$ state. Each line corresponds to the DEF for one Z-ancilla qubit. Here, the value of the first and last round is apparently lower than the others. This is a result of skipping idle operation which applies to all measurements except the last one. Since the first set of data is copied directly from the measurement without comparing with a physical data, the influence of an idle operation is also reduced. The curves in the middle are flat across different rounds, with slight lifting in the trend. It relates to leakage into a state out of the computational basis. Since the effect is minor, it indicates a low leakage error rate in our system. The average frequency for an error is about 37\%. We contribute this high rate to our long measurement duration. 
Figure~\ref{fig_4_def}(b) describes the same quantity for the logical $|-_L\rangle$ state. The experiment is done similarly except that we place a $R_Y(-\pi/2)$ gate in the beginning and $R_Y(-\pi/2)$ gate in the end of the circuit to each data qubit to produce 
$|-_L\rangle$ state after the first cycle, and measure all the data qubits in X basis after the final cycle.  We see similar result as we did for the logical $|0_L\rangle$ state, as expected.

The power of a surface code lies beyond it capability to detect errors. The pattern of the occurrence of these anomalies shed light on the errors happening on the data qubits. A powerful tool to visualize the correlation among each detection events is the correlation matrix. The correlation matrix describes the likelihood of observing a detection event at ancilla qubit $Dj$ at round $Rj$ given a error occurs at site $Di$ at round $Ri$, which can be written as
\begin{align}\label{eq:2}
p_{ij} = \frac{\langle p_{DiRi} p_{DjRj}\rangle - \langle p_{DiRi}\rangle \langle p_{DjRj}\rangle}{\sqrt{\langle \left(p_{DiRi}-\langle p_{DiRi} \rangle \right)^2\rangle \langle\left(p_{DiRi}-\langle p_{DiRi} \rangle \right)^2\rangle}}-\delta_{ij}
\end{align}
with $\delta_{ij}$ being the Kronecker delta with $\delta_{ij} = 1$ if $i=j$ and 0 otherwise.

The measurement we have made and the resulting syndromes are converted into a correlation matrix as shown in Fig.~\ref{fig_5_mat}(a) for logical $|0_L\rangle$ state and Fig.~\ref{fig_5_mat}(b) for logical $|-_L\rangle$ state.  The main diagonal blocks describe simultaneous correlation of detections among various ancilla qubits. The patterns it presents are consistent with the topology of the surface code and can be explained with potential crosstalks between neighbouring gates. The blocks away from the main  diagonal by a distance of one or two correspond to detection events one or two QEC cycles apart. The peaks in detection correlation in these blocks are due to error cumulation along the cycles since no reset operations are taken between two surface code cycles. As to the patterns further away, they may be explained by leakage to state out of the computational basis. Here, we have truncated the correlation matrix at $p=0$. The negative correlations are ignored which comes from us ignoring higher order correlations.

\begin{figure}[tbp]
\begin{center}
\includegraphics[width=1.0\linewidth]{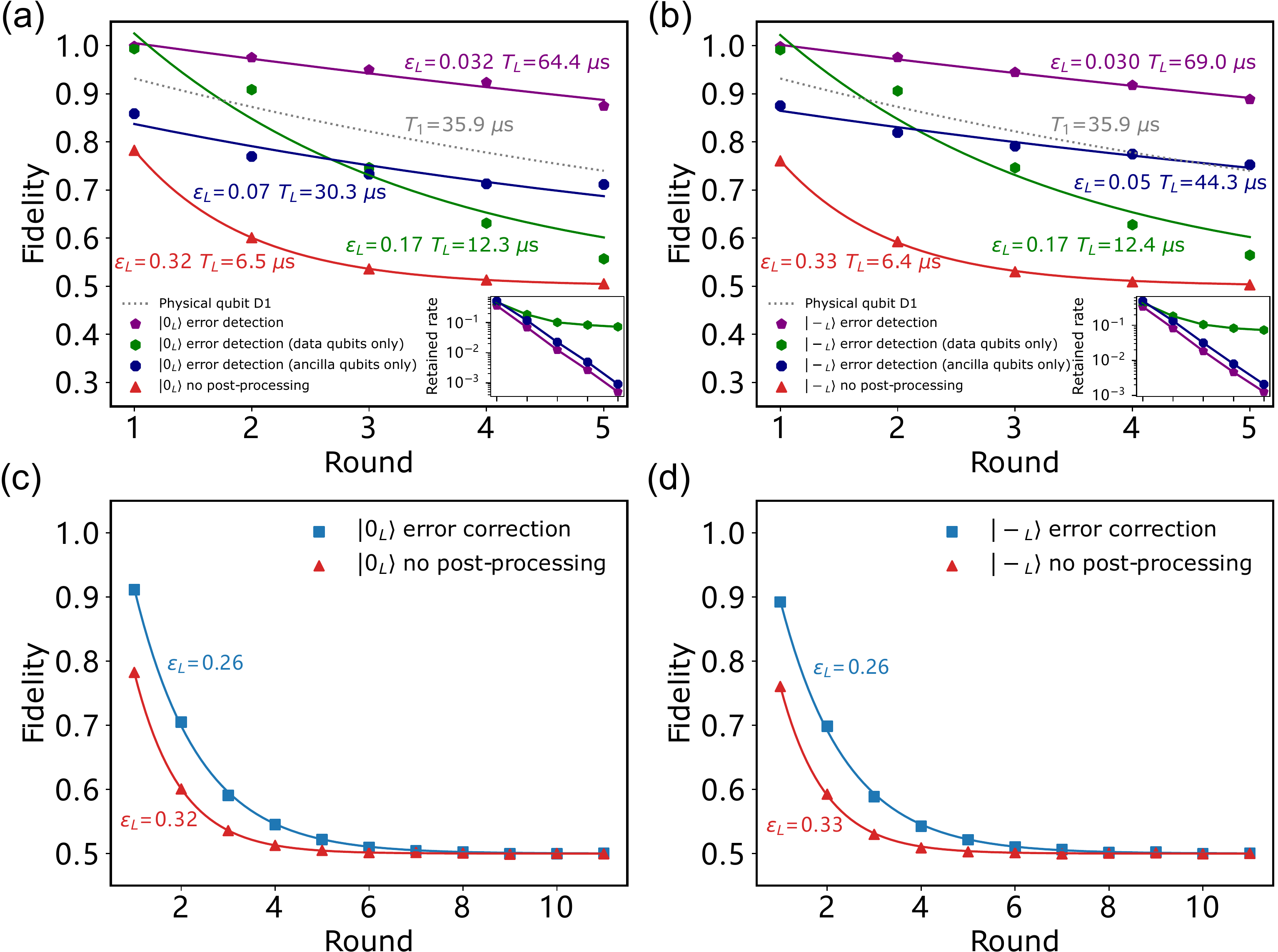}
\end{center}
\caption{\textbf{Results of error detection and error correction.} \textbf{(a)} Fidelity of the post-selected logical $|0_L\rangle$ state by error detection as function of clock cycles, i.e. the portion of samples that retain the logical state through some clock cycles. Various lines correspond to different post-selection schemes. The purple(pentagon) lines are obtained by discarding all data with error detected from the measurements of either data qubits or ancilla qubits. The green(hexagon) line corresponds to the dropping scheme based on data qubits only while the navy (octagon) line is only affected by the measurement of ancilla qubits. The red(triangle) line describes the result with no post selection. The dotted line depicts the prediction based on relaxation time $T_1$ of the best physical qubit among all used physical qubits. Logical error rates $\epsilon_L$ are extracted from the curve and logical fidelities $T_L$ are calculated. These values are listed by each line. Inset describes the retained rate for the three post-processing schemes as a function of round. \textbf{(b)} Results for the post-selected logical $|-_L\rangle$ state by error detection. \textbf{(c)} Fidelity of logical $|0_L\rangle$ state with the number of surface code cycles with error correction (blue line with square) or without (red line with triangular). \textbf{(d)} Same quantity for the logical $|-_L\rangle$ state after error correction.}
\label{fig_3_error_rate}
\end{figure}

Once an error has been detected, we can post-process the data and reduce the effect it brings to our system. One direct method is to drop erroneous data all in all. This simple treatment can greatly improve the fidelity of the logical state stored in our system by sacrificing efficiency of an data set. Here fidelity refers to the fraction of logical states that is intact after a certain clock cycles. The results are shown in Fig.~\ref{fig_3_error_rate}(a)-(b). There are five lines in each graph. The purple line is obtained as described before. That is to say, the fidelity is calculated using data where no error is detected. Here the green line describes the situation where we only drop a data if the final measurement of the data qubits implies an error. The navy line corresponds to the case where detection are based on measurement on ancilla qubits only. The red lines is obtained with all data without any post-processing. We have also include a grey dotted line describing the calculated fidelity based on the maximum physical relaxation time $T_1$ of participating physical qubits. By comparing these lines, we can see that both the measurement on data qubits and that on the ancilla qubit helps detecting errors and neither of them is capable of detecting all instances of error by itself. The inset describes the retained rate as a function of QEC round. Due to exponential drop of the retained rate, we did not exercise analysis for round greater than 5 with the number of shots for each experiment set to 480,000.

We derived the logical error rate by fitting the curves with the expression used in \cite{o2017density}
\begin{align}
\mathcal{F}_L(k) = \frac{1}{2} \left(1+(1-2 \epsilon_L)^{k-k_0}\right)
\end{align}
Here $k_0$ and $\epsilon_L$ relates to shift in round index and logical error rates which are to be fit. The fitted logical error rate are shown by each line. We also calculate the coherent time for the logical state using the following formula modified from \cite{o2017density} by consider decoherence from $|0\rangle$ and $|1\rangle$ state.
\begin{align}
T_L = \frac{\tau_{\text{cycle}}}{2\epsilon_L}
\end{align}
where $\tau_{\text{cycle}}$ is the wall-clock time for each cycle which can be calculated from the duration of single-qubit gate $\tau_{1Q}$, two-qubit gate $\tau_{2Q}$,
and measurement with waiting time for photon depletion $\tau_{M+D}$. To mitigate the distortion in the Z-pulse signal of a CZ gate between layer B and C, we insert an idle operation in-between for the same duration of a single-qubit gate, resulting a total duration of $5\tau_{1Q} + 4\tau_{2Q}+\tau_{M+D} = 4.153 \mu s$. Plug in this value, together as the error rate, we obtain the logical coherent time $T_L$ for each post-processing scheme. The result as listed by each line in Fig.~\ref{fig_3_error_rate}(a)(b). As can be seen from the results, after dropping instances errors detected in data qubits or measurement qubits, the lifetime of post-selected logial qubit $|0_L\rangle$ is $64.4 \mu s$, which exceeds the lifetime, $35.9\mu s$, of the best physical qubit among all used physical qubits.

Aside from dropping any corrupted data, a more efficient and sophisticated approach can be taken. That is to analyse the detection event, figure out when and where an error could possibly occur and propose a correction accordingly. One of the most widely used algorithm is the minimum weight perfect matching~\cite{higgott2021pymatching,fowler2012towards,kolmogorov2009blossom}. With the help of it, one can improve the fidelity of a logical state without losing a huge portion of the data generated. Figure~\ref{fig_3_error_rate} (c)-(d) show the fidelity of logical $|0_L\rangle$ and $|-_L\rangle$ state. The Red line describes the fidelity without an error correction while the blue line is obtained after all possible error corrections are made. The fitting results show that the logical errors ${\varepsilon _L}$ of $|0_L\rangle$ state and $|-_L\rangle$ state are reduced by $19\%$ and $21\%$, respectively, indicating the efficacy of the error correction algorithm we adopt. So far, the error rate is still higher compared to that of the physical error rate. We expect with number will be reduced with a longer code distance and shorter measurement time.

\section{CONCLUSION AND OUTLOOK}

Our experiment expands the surface code experiment's capabilities beyond error detection to include error correction. This 17-qubit surface code, in principle, already enables us to create a single fault-tolerant surface code memory. Future work will concentrate on realizing larger-scale surface codes, to achieve the important goal of suppressing the logical error rate as the code distance increases. This necessitates further improvements to the quantum computing system's performance, such as the number and quality of qubits, the fidelity of quantum gate operations, and rapid feedback of digital electronics.

\begin{acknowledgments}
The authors thank the USTC Center for Micro- and Nanoscale Research and Fabrication for supporting the sample fabrication. The authors also thank QuantumCTek Co., Ltd., for supporting the fabrication and the maintenance of room-temperature electronics.
\textbf{Funding:}
This research was supported by the National Key R\&D Program of China, Grant 2017YFA0304300, the Chinese Academy of Sciences, Anhui Initiative in Quantum Information Technologies, Technology Committee of Shanghai Municipality, National Science Foundation of China (Grants No. 11905217, No. 11774326),  Natural Science Foundation of Shanghai (Grant No. 19ZR1462700), and Key-Area Research and Development Program of Guangdong Provice (Grant No.2020B0303030001). H.-L. H. acknowledges support from the Youth Talent Lifting Project (Grant No. 2020-JCJQ-QT-030), National Natural Science Foundation of China (Grants No. 11905294), China Postdoctoral Science Foundation, and the Open Research Fund from State Key Laboratory of High Performance Computing of China (Grant No. 201901-01).

\textit{Note added}. Recently, we became aware of a similar work
by Sebastian $et$ $al.$~\cite{krinner2021realizing}, which was carried out independently.
\end{acknowledgments}

\bibliographystyle{apsrev4-1}
\bibliography{references}

\end{document}


\title{Supplemental Material for \\ ``Realization of an Error-Correcting Surface Code with Superconducting Qubits''}

\author{Youwei Zhao}
\thanks{These three authors contributed equally}
\author{Yangsen Ye}
\thanks{These three authors contributed equally}
\author{He-Liang Huang}
\thanks{These three authors contributed equally}
\author{Yiming Zhang}
\author{Dachao Wu}
\author{Huijie Guan}
\author{Qingling Zhu}
\author{Zuolin Wei}
\author{Tan He}
\author{Sirui Cao}
\author{Fusheng Chen}
\author{Tung-Hsun Chung}
\author{Hui Deng}
\author{Daojin Fan}
\author{Ming Gong}
\author{Cheng Guo}
\author{Shaojun Guo}
\author{Lianchen Han}
\author{Na Li}
\author{Shaowei Li}
\author{Yuan Li}
\author{Futian Liang}
\author{Jin Lin}
\author{Haoran Qian}
\author{Hao Rong}
\author{Hong Su}
\author{Lihua Sun}
\author{Shiyu Wang}
\author{Yulin Wu}
\author{Yu Xu}
\author{Chong Ying}
\author{Jiale Yu}
\author{Chen Zha}
\author{Kaili Zhang}
\author{Yong-Heng Huo}
\author{Chao-Yang Lu}
\author{Cheng-Zhi Peng}
\author{Xiaobo Zhu}
\author{Jian-Wei Pan}
\affiliation{Department of Modern Physics, University of Science and Technology of China, Hefei 230026, China}
\affiliation{Shanghai Branch, CAS Center for Excellence in Quantum Information and Quantum Physics, University of Science and Technology of China, Shanghai 201315, China}
\affiliation{Shanghai Research Center for Quantum Sciences, Shanghai 201315, China}

\date{\today}

\maketitle

\setcounter{section}{0}
\renewcommand{\thefigure}{S\arabic{figure}}	
\renewcommand{\thetable}{S\arabic{table}}	
\renewcommand{\theequation}{S\arabic{equation}}	
\setcounter{figure}{0}
\setcounter{table}{0}
\setcounter{equation}{0}

\section{Quantum processor and Experimental Wiring}
Our device is the same superconducting quantum processor in
Ref~\cite{zhu2021quantum} which consists of 66 Transmon
qubits~\cite{koch:042319} and 110 capacitive tunable couplers.  17
qubits and 24 couplers are choosed to perform our experiment which is shown
in Fig.1 in maintext. Data qubits and measure qubits are connected through
couplers which is more convinient for expanding our experiment to larger scale,
like $d=5$ repetition code. Each qubit has an inductively coupled control line
for both microwave control (XY) and flux control (Z) and each coupler has
an control line for flux control (Z). For performing parallel CZ gates,
each qubit is designed to have four arms to capactively couple to its four
nearest couplers and four nearset qubits~\cite{YangsenYe2021,sung2020}.
$\lambda/4$
readout resonators  coupled to corresponding $\lambda/2$ bandpass
filters~\cite{PhysRevLett.112.190504, zhu2021observation} are designed for
fast and accurate readout. Data qubits and measurement qubits share
different bandpass filters which lower the crosstalk of readout pulse
between them.
The experimental wiring setup and room temperature electronics are the same
in Ref.~\cite{zhu2021quantum}

\section{Basic calibration on 17 qubits and 24 couplers}
The basic calibration steps of the processor are the same with
\textit{Zuchongzhi} 2.0 including 17 qubits, 24 couplers, 17 readout resonators
and 7 JPAs are involved in this process. One can see more details about the
procedure in Ref~\cite{wu2021strong}.

\section{Fine calibration on 17 qubits and 24 couplers}
\subsection{Single-Qubit Gate Calibration and Readout}
For realizing high-fidelity parallel single qubit gates, we use the following
steps:

Step 1. Fine measurement of $T_1$ in large frequency range and XY crosstalk.

Step 2. Calculating qubits frequencies distribution with our classical
optimizer.

Step 3. Adjusting frequencies of couplers to minimize the effective coupling
strengths between nearest two qubits.

Step 4. Testing parallel single-qubit gates fidelities and adjust frequencies
to avoid two-level-systems.

The $T_1$ performance may change significantly during the experiment so we
repeat Step 3 and Step 4 several times to decrease the influence of
decoherence. Due to our control method of the system, we then calibrate the
square pulses of the qubit flux control lines to mitigate frequencies shift of
qubits~\cite{zhu2021quantum}. We then measure the single-qubit gate fidelity
simultaneously by cross-entropy benchmarking(XEB)~\cite{arute2019quantum}. All
selected qubits are placed at idle frequencies during measurement. Readout
fidelity of each qubit is measured after qubits are prepared at
$|00,...,0\rangle$ or $|11,...,1\rangle$. The device parameters of
single-qubit  are summarized in Table \ref{tableSummary} and gate errors are
average errors.

\begin{table*}[htb!]
	\centering
	\begin{tabular}{c| c c c c c c c c c c}
		\hline
		\hline
		Parameters& X1& Z1& X2& Z2& Z3& X3& Z4& X4& &Average\\
		\hline
		\hline
		Qubit maximum frequency, $\omega_{\textrm{q}}^{\textrm{max}}/2\pi$
		(GHz), & 5.079& 5.103& 4.990& 5.118& 5.146& 5.034& 5.079& 5.226& & 5.097
\\
		\hline
		Qubit idle frequency, $\omega_{\textrm{q}}/2\pi$ (GHz)& 4.876& 5.060& 4.949& 5.115& 4.950& 4.907& 5.017& 5.168& & 5.005
\\
		\hline
		Readout drive frequency, $\omega_{\textrm{r}}/2\pi$ (GHz)& 6.352& 6.467& 6.412& 6.352& 6.410& 6.347& 6.288& 6.409& & 6.380
\\
		\hline
		Qubit anharmonicity, $U/2\pi $ (MHz)& -248& -242& -242& -236& -243& -244& -248& -248& & -244
\\
		\hline
		Energy relaxation time $T_1$ at idle frequency ($\mu$s)& 30.3& 18.5& 26.9& 26.2& 20.9& 22.3& 22.1& 21.1& & 23.5
\\
		\hline
		Echo decay time $T_2$ at idle frequency ($\mu$s)& 7.1& 5.9& 8.6& 9.8& 9.9& 5.7& 4.7& 3.2& & 6.9
\\
		\hline
		Ramsey decay time $T_2^*$ at idle frequency ($\mu$s)& 3.4& 3.3& 5.7& 5.4& 6.3& 4.7& 4.3& 1.3& & 4.3
\\
		\hline
		Readout	$F_{00}$ (\%)& 98.6& 98.5& 96.1& 98.5& 96.6& 98.6& 97.2& 97.8& & 97.7
\\
		\hline
		Readout $F_{11}$ (\%)& 94.0& 95.0& 91.7& 93.4& 91.4& 90.7& 91.3& 93.9& & 92.7

		\\
		\hline
		Readout linewidth, $\kappa$/2$\pi$ (MHz)& 1.11& 0.47& 1.08& 1.53& 1.87& 0.85& 0.51& 0.96& & 1.05
\\
		\hline
		1Q XEB $e_{1}$ (\%))& 0.060& 0.121& 0.116& 0.152& 0.124& 0.125& 0.076& 0.102& & 0.110
\\
		\hline
		1Q XEB $e_{1}$ purity (\%)& 0.058& 0.102& 0.111& 0.128& 0.104& 0.100& 0.066& 0.087& & 0.094
\\
		\hline
		\hline
		Parameters& D1& D2& D3& D4& D5& D6& D7& D8& D9& Average\\
		\hline
		\hline
		Qubit maximum frequency, $\omega_{\textrm{q}}^{\textrm{max}}/2\pi$
		(GHz) & 5.017& 4.831& 4.846& 5.292& 5.184& 5.108& 5.244& 5.261& 5.185& 5.108
\\
		\hline
		Qubit idle frequency, $\omega_{\textrm{q}}/2\pi$ (GHz)& 4.852& 4.782& 4.744& 4.780& 4.824& 4.726& 4.827& 4.789& 4.831& 4.795
 \\
		\hline
		Readout drive frequency, $\omega_{\textrm{r}}/2\pi$ (GHz)& 6.382& 6.437& 6.491& 6.380& 6.435& 6.488& 6.374& 6.428& 6.481& 6.433
 \\
		\hline
		Qubit anharmonicity, $U/2\pi $ (MHz)& -246& -247& -249& -246& -247& -249& -248& -246& -252& -248
  \\
		\hline
		Energy relaxation time $T_1$ at idle frequency ($\mu$s)& 35.9& 35.8& 26.6& 22.0& 30.2& 31.4& 24.5& 27.8& 21.5& 28.4
\\
		\hline
		Echo decay time $T_2$ at idle frequency ($\mu$s)& 4.9& 4.5& 3.8& 6.3& 8.4& 3.6& 4.1& 4.9& 7.0& 5.3
\\
		\hline
		Ramsey decay time $T_2^*$ at idle frequency ($\mu$s)& 2.1& 1.8& 2.6& 3.2& 6.2& 1.6& 1.3& 2.2& 5.2& 2.9
\\
		\hline
		Readout	$F_{00}$ (\%)& 97.6& 96.1& 97.5& 97.7& 97.8& 98.0& 97.8& 98.3& 98.8& 97.7
 \\
		\hline
		Readout $F_{11}$ (\%)& 92.3& 90.1& 94.1& 92.7& 93.0& 92.9& 93.9& 93.9& 92.8& 92.9
\\
		\hline
		Readout linewidth, $\kappa$/2$\pi$ (MHz)& 0.99& 1.16& 0.33& 0.96& 0.47& 0.36& 1.22& 0.77& 0.59& 0.76
 \\
		\hline
		1Q XEB $e_{1}$ (\%)& 0.087& 0.117& 0.072& 0.102& 0.077& 0.063& 0.074& 0.084& 0.118& 0.088
\\
		\hline
		1Q XEB $e_{1}$ purity (\%)& 0.073& 0.121& 0.066& 0.096& 0.093& 0.054& 0.065& 0.082& 0.087& 0.082
\\
		\hline
		\hline
	\end{tabular}
	\caption{\textbf{Summary of single-qubit parameters.}}
	\label{tableSummary}
\end{table*}

\begin{table*}[htb!]
	\centering
	\begin{tabular}{c| c  c c c c c c}
		\hline
		\hline
		Apattern& D1-Z1& D2-X2& D3-Z2& D5-Z3& D6-X3& D8-X4& Average\\
		\hline
		2Q XEB $e_{2}$ (\%)& 0.70& 1.69& 0.77& 0.99& 1.30& 0.70& 1.02
\\
		\hline
		\hline
		Bpattern& D1-X2& Z1-D4& Z2-D6& D5-X3& Z3-D8& D7-X4& Average\\
		\hline
		2Q XEB $e_{2}$ (\%)& 2.40& 0.75& 0.87& 1.14& 0.62& 0.55& 1.05
\\
		\hline
		\hline
		Cpattern& X1-D3& D2-Z2& X2-D5& D4-Z3& D6-Z4& X3-D9& Average\\
		\hline
		2Q XEB $e_{2}$ (\%)&1.58& 0.85& 0.71& 1.30& 1.02& 1.37& 1.14
\\
		\hline
		\hline
		Dpattern& X1-D2& X2-D4& Z2-D5& Z3-D7& X3-D8& Z4-D9& Average\\
		\hline
		2Q XEB $e_{2}$ (\%)& 1.20& 1.46& 0.80& 0.68& 0.93& 0.47& 0.92
\\
		\hline
		\hline
	\end{tabular}
	\caption{\textbf{Summary of CZ gates parameters.}}
	\label{tabletwoqubitgatesSummary}
\end{table*}

\subsection{CZ Gate Calibration}
CZ gates in our experiment are realized by tuning the frequencies of qubits to
make $|11\rangle$ states in resonance with $|02\rangle$ states and carefully
tuning the frequencies of couplers to open the coupling
strengths~\cite{YangsenYe2021,sung2020}.
For realizing high-fidelity parallel CZ gates, we use the following
steps:

Step 1. Fine measurement of $T_1$ in large frequency range.

Step 2. Calculating qubits working frequencies distribution when performing
parallel CZ gates with our classical optimizer for different patterns.

Step 3. Testing parallel CZ gates fidelities and adjust working frequencies to
avoid two-level-systems.

In our classical optimizer to optimize the working frequencies of CZ gates, we
consider several error estimation models, including decoherence error, residual
coupling strength error and flux control pulse distortion error. To lower the
two-level-systems'influence, we adjust the working frequencies distribution by
scanning the relationship between working frequencies and the speckle purity
benchmarking(SPB)~\cite{arute2019quantum}.  We then measure the CZ gates
fidelities by cross-entropy benchmarking(XEB)~\cite{arute2019quantum} of four
patterns. The device parameters of CZ gates are summarized in Table
\ref{tabletwoqubitgatesSummary} and gate errors are average errors.

\begin{figure*}[!htbp]
\begin{center}
\includegraphics[width=1\linewidth]{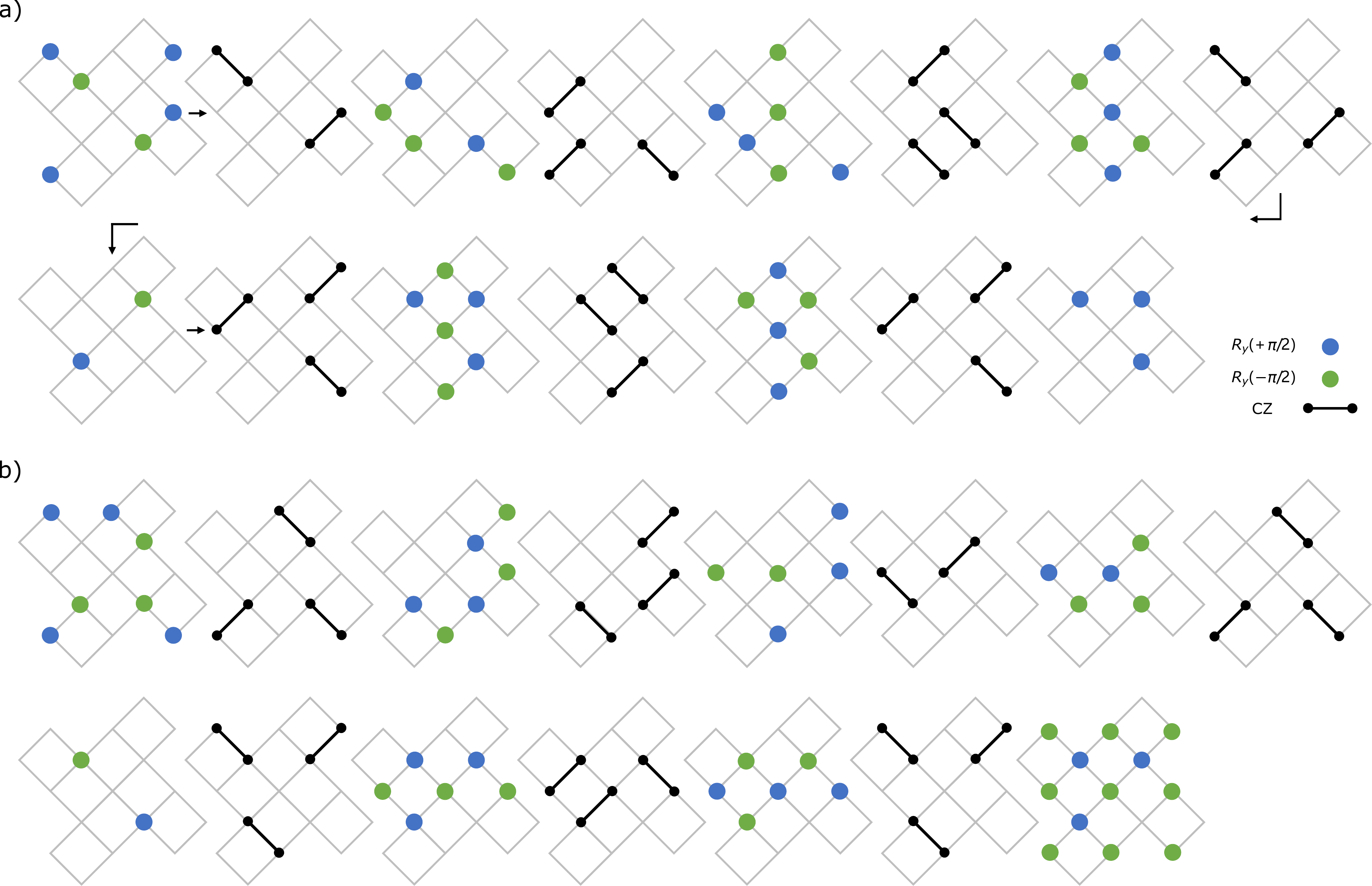}
\end{center}
\caption{\textbf{a)} Preparation of the $|0_L\rangle$ without measurement, starting from $|0\rangle^{\otimes 17}$, ending with 9 data qubits in logical state, 8 ancilla qubits in $|0\rangle^{\otimes 8}$. \textbf{b)} Preparation of the $|-_L\rangle$ without measurement.}
\label{sfig_7stepinit}
\end{figure*}

\begin{figure}[!htbp]
\begin{center}
\includegraphics[width=0.9\linewidth]{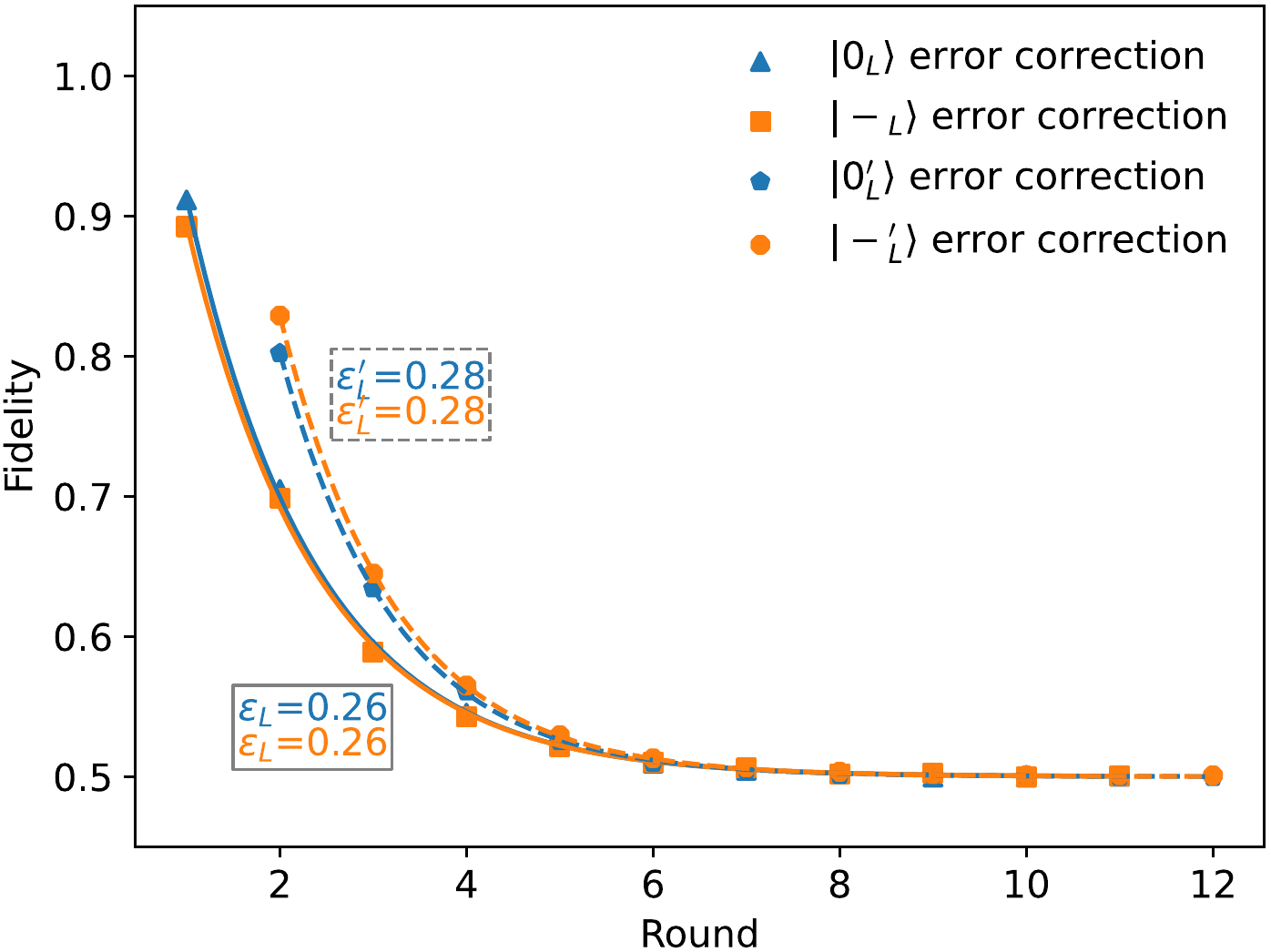}
\end{center}
\caption{Fidelity of logical $|0_L\rangle$ and $|-_L\rangle$ state prepared with one  surface code cycle (solid line) and those generated with steps described in Figure~\ref{sfig_7stepinit}(dashed line). $\epsilon_L$ are logical error rates extracted from the corresponding curve.}
\label{sfig_error_rate}
\end{figure}

\begin{figure}[!htbp]
\begin{center}
\includegraphics[width=0.95\linewidth]{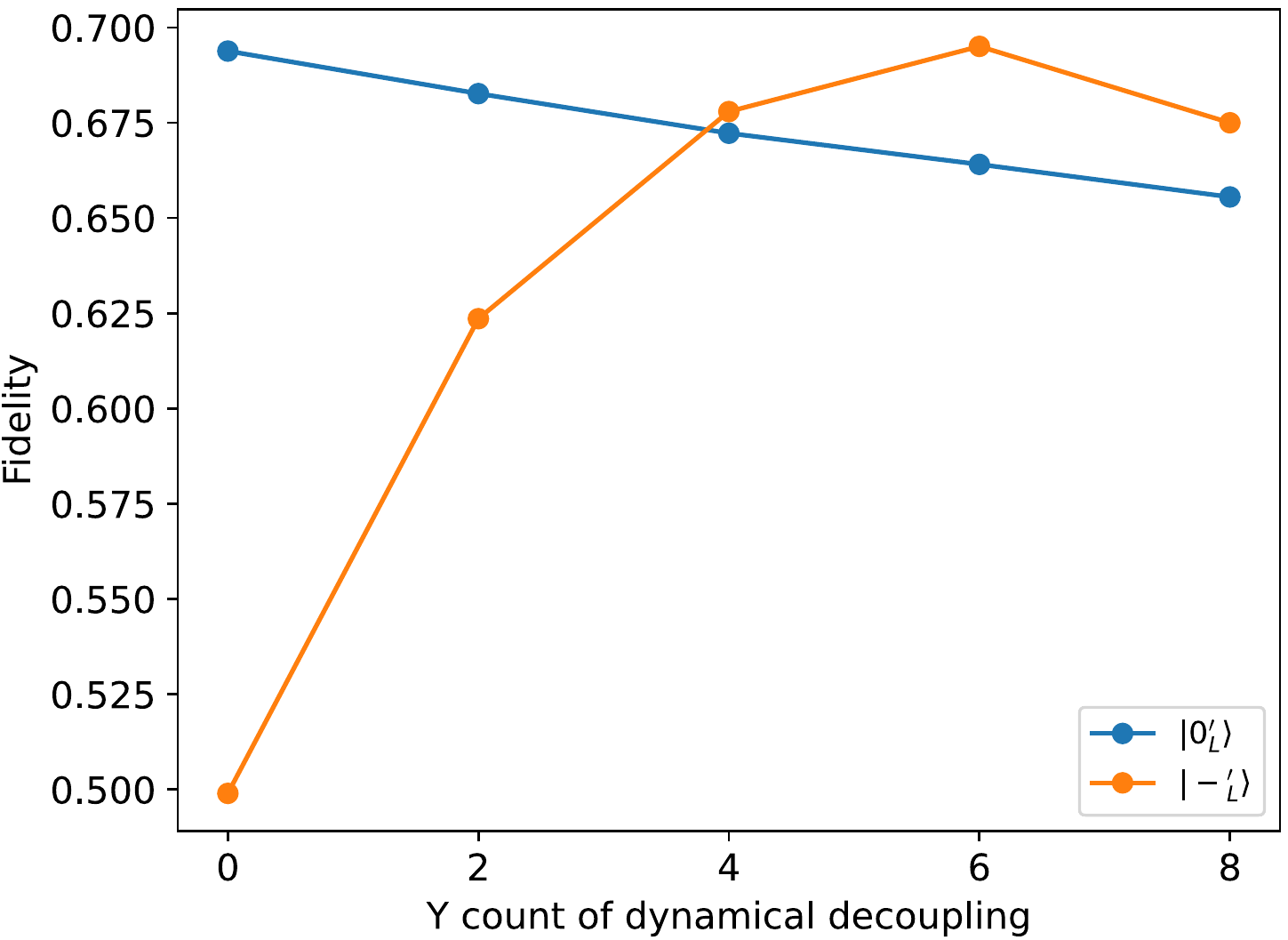}
\end{center}
\caption{Fidelity of logical state with different Y gate counts in dynamical decoupling operator. 
Logical states are prepared with approach described in the supplementary material
. Blue line is for logical $|0\rangle$ state and orange line is for logical$|-\rangle$.}
\label{sfig_dd}
\end{figure}

\section{Logical state preparation without measurement}
In the main text, we initialize a logical $|0_L\rangle$ and $|-_L\rangle$ state with one surface code cycle as presented in Figure 1(b). Here we describe another scheme for preparing logical states inspired by the method used in Ref.~\cite{doi:10.1126/science.abi8378}. The steps are shown in Fig.~\ref{sfig_7stepinit}. After 15 steps with either single-qubit or two-qubit gates, we obtain the exact target logical state. We also carry out error correction experiments based on states prepared in this way and compare the trend of logic state fidelity with that of the conventional method discussed in the main text. The result are shown in Fig.~\ref{sfig_error_rate}. As the plot shows, the results are close to each other. This provides another method that is friendly to systems without repetitive measurement.

\section{Y-gate Count Optimization in Dynamical Decoupling}
Dynamical decoupling is implemented by applying a Y rotation to a qubit flips the state relative to the plane of equator on a Bloch Sphere and reverses the drifting direction for some time. To find out the optimal structure of a dynamical decoupling operator, we test the influence of the Y-gate count on the logical fidelity with various number of Y gates. For each experiment, a logical state, be it $|0_L\rangle$ or $|-_L\rangle$, is created using the method discussed immediately above. Then
a measurement is taken on the ancilla qubits and dynamical decoupling operators are applied to the data qubits. Here we experiment with different number of Y gates and extract the fidelity of the logical state from the data qubit measurement. The result are shown in Fig.~\ref{sfig_dd}. With the data available, we set the optimal dynamical decoupling configuration to be with 6 Y gates.

\bibliography{references}